\documentclass[aps,prl,groupedaddress]{revtex4}
\usepackage{graphicx}
\usepackage{amsmath}
\def\un#1{\underline{#1}}
\begin{document}
%JOEL:
\large \baselineskip=24pt

\title{THE BOLTZMANN ENTROPY FOR DENSE FLUIDS NOT IN LOCAL EQUILIBRIUM}
\author{P.L. Garrido$^{(1,2)}$, S. Goldstein$^{(2)}$ and J.L. Lebowitz$^{(2)}$}
%JOEL:
%\address{Departamento de E.M. y F{\'\i}sica de la Materia. Universidad de Granada.
%E-18071 Granada. Spain\\
%Department of Mathematics and Physics. Rutgers University. NJ 08903, USA} \maketitle
%END JOEL
%PRL:
\affiliation{(1) Depart. de E.M. y F{\'\i}sica de la Materia. Universidad de Granada.
E-18071 Granada. Spain\\
(2) Department of Mathematics and Physics. Rutgers University. NJ 08903, USA}
%%%END PRL

\begin{abstract}
We investigate, via computer simulations, the time evolution of the (Boltzmann) entropy of a dense
fluid not in local equilibrium. The macrovariables $M$ describing the system are the (empirical)
particle density $f=\{f(\un{x},\un{v})\}$ and the total energy $E$.  We find that $S(f_t,E)$ is
monotone increasing in time even when its kinetic part is decreasing. We argue that for isolated
Hamiltonian systems  monotonicity of $S(M_t) = S(M_{X_t})$ should hold generally for ``typical''
(the overwhelming majority of) initial microstates (phase-points) $X_0$ belonging to the initial
macrostate $M_0$, satisfying $M_{X_0} = M_0$.  This is a direct consequence of Liouville's theorem
when $M_t$ evolves autonomously.
% e.g.\ for hydrodynamical variables for systems in local
%equilibrium (obeying Euler or Navier-Stokes equations), and for the density $f_t$ for dilute gases
%(Boltzmann equation) and moderately dense hard-spheres (modified Enskog equation).
\end{abstract}
%PRL:
\maketitle

\section{INTRODUCTION}

In 1854 Clausius introduced the notion of the entropy of a macroscopic system (defined up to
additive constants) in a state of thermal equilibrium \cite{Jaynes1}.  Not long afterwards
Boltzmann gave a microscopic definition of the entropy $S(M)$ of a general macroscopic system in a
macrostate defined by values of macrovariables $M$: $M=M_X$ represents suitable ``coarse-grained''
functions of the system's microstate, given by a point $X$ in the $2Nd$-dimensional phase space
$\Gamma$ for a $d$-dimensional classical system containing $N$ particles. $S(M)$ is (up to
constants) equal to the log of the volume $|\Gamma_M|$ of the phase-space region $\Gamma_{M}$,
defined by the macrostate $M$, i.e.\ containing all phase points (or
microstates)
giving rise to this macrostate \cite{Jaynes2}. $S(M)$ agrees with the Clausius entropy for systems
in equilibrium when $M = M_{eq}$ is just the total energy $E$, of $N$
particles in a volume $V$. The fact that
$|\Gamma_{M_{eq}}|$ is exponential in the number of molecules in the system explains the origin of
the second law in the microscopic dynamics: when a constraint is lifted in a system in equilibrium,
thereby affording access to a new equilibrium macrostate $M^\prime_{eq}$ of larger entropy per
molecule than the original $M_{eq}$, the overwhelming majority of phase points in $\Gamma_{M_{eq}}$
will find themselves eventually in $\Gamma_{M^\prime_{eq}}$, since for a macroscopic system
$|\Gamma_{M^\prime_{eq}}|$ is enormously larger than
$|\Gamma_{M_{eq}}|$ as well as the volume of the union of
all nonequilibrium macrostates without the constraint \cite{Jaynes2} \cite{Boltz}.

Boltzmann's interpretation of entropy naturally extends the second law to nonequilibrium
macroscopic systems:  a (significant) violation of the second law will not occur provided
% which are: (i) characterized by
%macrovariables for which $S(M)$ like $S(M_{eq})$ is extensive in the
%size of the system and (ii)
the microstate $X$ of a physical system prepared in, or evolved into,
the macrostate $M$ is typical (or at least not atypical) of points in
$\Gamma_M$ as far as the future evolution of $M$ is concerned.  Of
course to rigorously prove that $X$ should be (and remain) typical in
this sense is difficult.  It was pointed out, however, in \cite{Resib}
that a sufficient condition for this is that the evolution of $M_t$ be
given by an autonomous deterministic law, i.e.\ the value of $M$ at
any time $t+\tau$ is determined by its value at time $t$,
$M_{t+\tau}=\Phi_\tau(M_t)$, for $\tau \geq 0$.  Such a law implies
that, with rare exception, $\phi_\tau \Gamma_{M_t} \subset
\Gamma_{M_{t+\tau}}$, i.e.\ the overwhelming majority of phase-points
(almost all in suitable limits) $X \in \Gamma_{M_t}$ will, when
evolved according to the microscopic evolution law $\phi_\tau$, be
found in $\Gamma_{M_{t+\tau}}$ for $\tau \geq 0$.  It follows then
from Liouville's theorem that (with insignificant error)
$S(M_{t+\tau}) \geq S(M_t)$.

The most common example of a deterministic macro-evolution occurs for
simple systems in local thermal equilibrium (LTE), described by the
macrovariables $M$ representing the locally conserved (particle,
momentum and energy) densities.  $M_t$ then satisfies deterministic
hydrodynamic-type equations, e.g.\ the Euler or Navier-Stokes
equations, and $S(M_t)$, which is given by an integral over the volume
$V$ of the equilibrium entropy density, see \cite{Alder}, is then
indeed monotone nondecreasing in time.

Such a deterministic evolution is however not necessary for the
monotonicity of $S(M_t)$.  There is an enormous disparity between the
(small) number of possible macrostates (which are always defined with
some macroscopic tolerance in terms of a relatively small number of
macrovariables) and the large number of microstates corresponding to
the possible values of the large number of microvariables (counted say
in terms of phase-space cells of volume $\bar h^{d N}$).  Even when
$M_0$ does not determine $M_t$, $S(M_{X_t})$ should be nondecreasing to
leading order in the size of the system for the overwhelming majority
of initial microstates $X_0$; see also \cite{Ehrenfest}.

The choice of macrovariables $M$ and the corresponding computation of $S(M)$ suitable for
describing systems not in LTE is a daunting task, especially  for complex systems such as polymeric
fluids, metals with memory, etc.   The first and still paradigmatic step in that
direction was taken by Boltzmann himself when he computed $S(f)$ for the macrovariables  $f =
\{f_X(\un{x},\un{v})\}$ corresponding to the coarse-grained (empirical) density in the  $2d$
dimensional $\mu$-space for a macroscopic system in a microstate $X$
\cite{Jaynes2},\cite{Goldstein}. He found that (up to constants)
\begin{equation}
S(f) = S_{gas}(f) \equiv -k\int_V d\un{x}\int_{{{R}}^d}d\un{v}f(\un{x},\un{v})\ln
f(\un{x},\un{v})\label{Bolt}
\end{equation}

We put the subscript ``gas'' on $S_{gas}(f)$ to emphasize that $f(\un{x}, \un{v})$ can be expected to
suffice for the adequate specification of the macrostate away from LTE only for a dilute gas, where
interactions between particles make a negligible contribution to the
energy of the system.  In fact it was for such a dilute gas that
Boltzmann derived a deterministic evolution equation for $f$ and
proved that the corresponding $f_t$ satisfies the ${\cal H}$-theorem,
${d \over dt} S_{\rm gas}(f_t) \geq 0$, \cite {Jaynes2}, \cite{Boltz},
\cite{Goldstein}.    For
dense fluids, specification of $f(\un{x},\un{v})$ is compatible with many different total energies
(including infinite ones for hard core interparticle potentials).  A simple analysis then shows
that the phase points $X$ of a system with specified energy $E$ which is below the maximal energy
compatible with $f$ will correspond to an exceedingly small minority of the phase points in
$\Gamma_f$, i.e.\ will be atypical of points in $\Gamma_f$. There is then no reason to expect for
such systems that $S_{gas}(f_{X_t})$ will increase as the system evolves in time according to its
energy conserving Hamiltonian dynamics. It is in fact easy to set up in dense fluids initial
macrostates $f_0$ such that for $X_0 \in \Gamma_{M_0}$
$S_{gas}(f_{X_t})$  will typically
 decrease in time when the fluid goes to equilibrium
\cite{Lebowitz1}.

It was argued in \cite{Resib} that if one includes in $M$, in addition to $f$, also the total
energy $E$, then the entropy $S(f_t,E)$ should be an increasing function of time, i.e. $S(f_t,E)$
should satisfy an ${\mathcal{H}}$-theorem for general systems, including dense fluids. It was also
noted there that the quantity shown by Resibois \cite{Rapaport} to satisfy an
${\mathcal{H}}$-theorem for $f_t$ evolving via the modified Enskog equation (expected to be
accurate for moderately dense hard sphere gases) is in fact the Boltzmann entropy $S(f_t,E)$ for a
system of hard spheres.

In this work we use molecular dynamics to investigate the time
evolution of $S(f_t,E)$ for dense fluids interacting with
Lennard-Jones and other types of pair potentials. We consider in
particular situations, such as those in \cite{Lebowitz1}, where
$S_{gas}(f_t)$, defined in (\ref{Bolt}), is expected to decrease. Our
simulations, which give a monotone increase of $S(f_t,E)$ when the
number of particles in the system is large, support the hypothesis
that the time evolution of a typical microstate in $\Gamma_{f,E}$ is
indeed such that $S(f_t,E)$ satisfies an ${\mathcal{H}}$-theorem.  We
also find evidence that $f_t$ itself evolves in a deterministic way,
with different microstates with the same $f_0$ give rise to the same
$f_t$, although no equation yielding this evolution is at present
known (at least to us) for general dense fluids.  This suggests
looking for an autonomous equation for $f_t$ (at a given $E$).  This
is exactly what is done in the heuristic derivations of the Boltzmann
and Enskog equations \cite{Lennard} and is discussed extensively in
the literature for various other systems, see \cite{11}. The validity
of the Boltzmann equation for dilute gases, i.e.\ for typical $X_0 \in
\Gamma_{f_0}$ in the Boltzmann-Grad limit, was justified rigorously at
least for short times, by Lanford \cite{Lanford}.

{\bf Formalism:} We consider a system of $N$ particles with unit mass in a box $V$. The microstate
is specified by $X=(\un{x}_1,\un{v}_1,\ldots ,\un{x}_N,\un{v}_N)$ and the dynamics is given by the
Hamiltonian
\begin{equation}
H(X)=\frac{1}{2}\sum_{i=1}^N\un{v}_i^2+\frac{1}{2}\sum_{i\neq j}\phi(\un{x}_i-\un{x}_j)
\label{Hamil}
\end{equation}
It follows from the structure of the classical phase space  $\Gamma$ that $S(f,E)$ can be written,
c.f. \cite{Resib}, as
\begin{equation}
S(f,E)=S^{(m)}(f)+S^{(c)}(n,\Phi_{total}). \label{entropy}
\end{equation}
Here $S^{(m)}$ ($S^{(c)}$) is the log of the momentum (configuration) space volume corresponding to
the macro-state $M=(f,E)$,
\begin{equation}
S^{(m)}(f)=S_{gas}(f)+k\int_V d\un{x}\, n(\un{x})\log n(\un{x})\label{dos}
\end{equation}
with $n(\un{x})=\int_{{R}^{d}}d\un{v}f(\un{x},\un{v})$, the spatial density, and
\begin{equation}
S^{(c)}(n,\Phi_{tot})=\sup_{\Phi}\int_V d\un{x}\, s^{(c)}(n(\un{x}),\Phi(\un{x}))\label{conf}
\end{equation}
where $s^{(c)}(n,\Phi)$ is the configurational entropy density of an
equilibrium system  with
Hamiltonian (\ref{Hamil}) having particle density $n$ and potential energy density $\Phi$. The
$\sup$ in eq. (\ref{conf}) is taken over all $\Phi(\un{x})$ such that
\begin{equation}
\int_Vd\un{x}\,\Phi(\un{x})=\Phi_{total}=E-\int_{V}d\un{x}\int_{R^d}d\un{v}f_t(\un{x},\un{v})
\frac{1}{2}\un{v}^2 \label{tres}
\end{equation}

Restricting ourselves to spatially uniform systems, $n=N/V$, $f=nh(\un{v})$, $\int_{{{R}^d}}d\un{v}
\un{v} h(\un{v})=0$, $\Phi=\Phi_{total}/V$, we find (see eqs. (39)-(41) in \cite{Goldstein}) that
\begin{equation}
\frac{d}{dt}S^{(c)}(n, \Phi_{\rm total})=\frac{1}{T_{\Phi}}\frac{d\Phi_{total}}{dt}
\label{entrevol}
\end{equation}
where $T_{\Phi}$ is the inverse of $\Phi_{eq}(T)$, the potential energy density of the equilibrium
system with Hamiltonian (\ref{Hamil}) at density $n$.

{\bf Simulations:} To check whether $S(f_t,E)$ as expressed in eqs. (\ref{entropy}) to
(\ref{entrevol}) satisfies an ${\cal H}$-theorem for dense fluids we have carried out simulations
on a two-dimensional system with density $n=0.5$ in a periodic box interacting with a cut-off
Lennard-Jones potential, $\phi(r)=\bar\phi(r)-\bar\phi(r_c)-(r-r_c)d\bar\phi(r)/dr\vert_{r=r_c}$
with $\bar\phi(r)=4[r^{-12}-r^{-6}]$ for $r\leq r_c=2.5$ and $\phi(r)=0$ otherwise. To obtain
$T_\Phi$ in (\ref{entrevol}) we first computed $\Phi_{eq}(T)$,
for such a system.   For these simulations, as well as for those described below, we used the
Verlet algorithm with time mesh $10^{-4}$.

\begin{figure}
\centerline{\includegraphics[width=0.5 \textwidth]{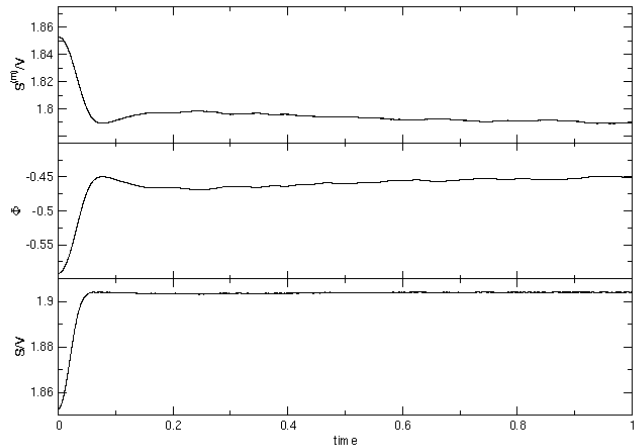}} \caption{ Evolution of $S^{(m)}/V$,
$\Phi$ and $S/V$ following the initial nonequilibrium state (see text). The particles interact with
a cut-off Lennard-Jones potential and $N=90000$. The total energy and the initial potential energy
densities are $e=0.6$ and $\Phi=-0.5917..$ respectively.}\label{evolLJm12}
\end{figure}

To carry out the time dependent simulations, we first let the system
reach an equilibrium state.  We then multiplied the speed of each
particle with an appropriate factor to obtain a state with too high a
kinetic energy, and let the system evolve freely to its new
equilibrium state. This means that the initial $f$ was a Maxwellian
with too high a temperature for the total energy---the case considered
in \cite{Lebowitz1}.  During the consequent evolution the velocity
distribution stayed isotropic and we computed $S^{(m)}$, as well as
the kinetic and potential energy densities, in time intervals of size
$0.001$. To compute $S^{(m)}$ we first find the speed distribution by
counting the number of particles having their speeds in each of the
$50$ equal intervals into which we divide the segment $[0,v_{max}]$
where $v_{max}$ is the maximum speed of any particle at time $t$. We
then get the configurational entropy $S^{(c)}$ by taking a numerical
derivative of $\Phi$ with respect to time and then carrying out a
numerical integration of the right side of (\ref{entrevol}). In Figure
\ref{evolLJm12} we show the time behavior of $S^{(m)}$, $\Phi_{\rm
total}$ and $S$ for such a system.

%\begin{figure}
%\centerline{\includegraphics[width=0.5\textwidth]{./figuras_prl/equilLJm12_prl.ps}}
%\caption{The equilibrium potential energy density vs. the equilibrium
%kinetic energy density for the Lennard Jones potential, $N=900$,$1600$
%and $90000$ particles with $n=0.5$. Error bars are smaller than the
%symbol sizes. The arrows show the typical process we follow to prepare
%the system and then evaluate its nonequilibrium entropy (see main
%text).}\label{equiLJ12}
%\end{figure}

\begin{figure}
\centerline{\includegraphics[width=0.5\textwidth]{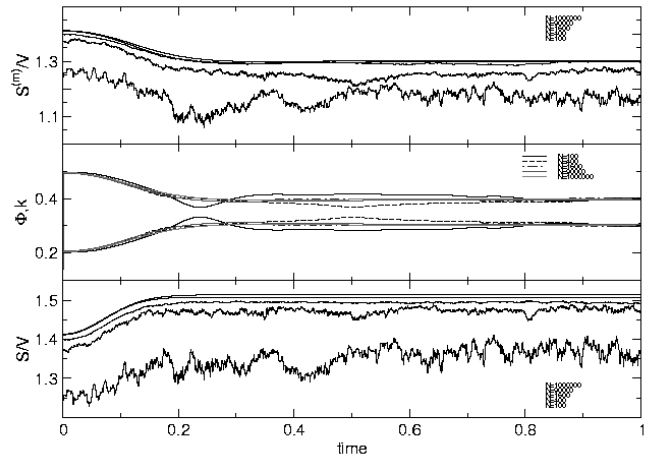}} \caption{Evolution of $S^{(m)}/V$, $k$,
$\Phi$ and $S/V$ for different size systems $N=100, 400, 1600, 90000 $ and $10^6$ .  The particles
interact with a cut-off  $r^{-6}$ potential. The total energy and the initial potential energy
densities are $e=0.7$ and $\Phi=0.2049..$ respectively.} \label{evolLJm6}
\end{figure}

We have carried out similar calculations for the truncated repulsive potential
$\phi(r)=\bar\phi(r)-\bar\phi(r_c)-(r-r_c)d\bar\phi(r)/dr\vert_{r=r_c}$ with $\bar\phi(r)=r^{-6}$
for $r\leq r_c=2.5$ and $\phi(r)=0$ otherwise. We show the results for different values of $N$ in
Fig. \ref{evolLJm6}. The fluctuations for small $N$ are clearly visible: their magnitude, once the
system has reached equilibrium, appear to scale as $N^{-1/2}$. There are also finite-size
corrections to the equilibrium time averages consistent with those expected from using a
micro-canonical ensemble.

\begin{figure}
\centerline{\includegraphics[width=0.5 \textwidth]{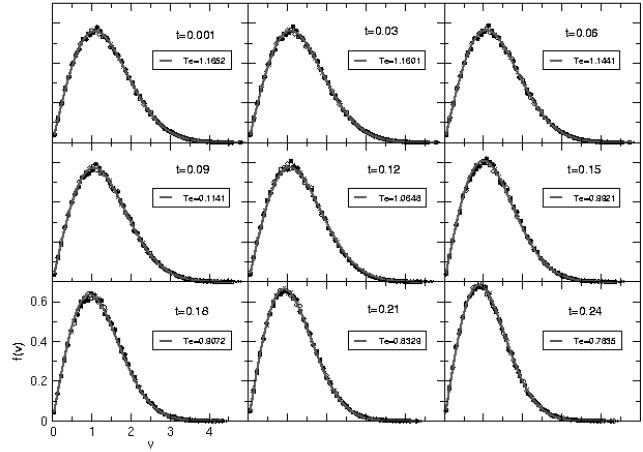}} \caption{Evolution of $f_t(v)$ during
a ($B$) to ($C$) process for three different microscopic configurations. The particles interact
with a truncated $r^{-6}$ potential and $N=90000$. The solid lines correspond to the fit of the
data to a Maxwellian distribution with the same kinetic energy}\label{3cond}
\end{figure}

\begin{figure}
\centerline{\includegraphics[width=0.5 \textwidth]{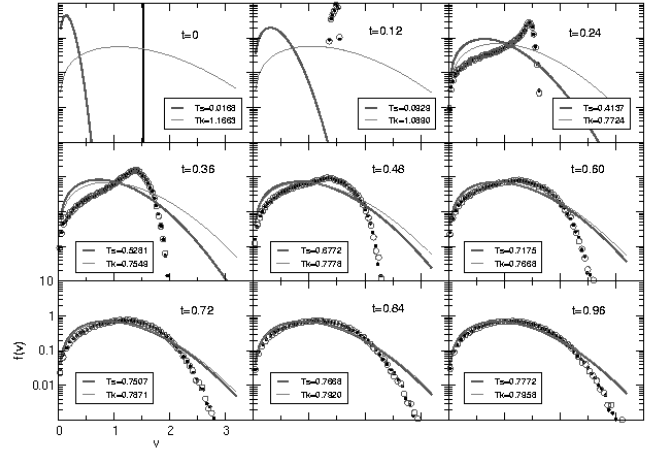}} \caption{Evolution of $f_t(v)$ for an
initial condition with all particles in a square lattice and equal speeds with random directions
for $n=0.5$, $e=0.7$, $\bar\phi(r)=r^{-6}$and $N=90000$. Full dots and empty circles are the values
of $f_t$ corresponding two different microscopic states respectively.   Gray lines correspond to
Maxwellian distributions with temperature $T$ obtained from the kinetic energy of the state.
}\label{fdist05}
\end{figure}

We also investigated for this system whether $f_t$ evolves deterministically by comparing $f_t$ for
different initial microstates, all having the same $f_0$ and $E$. The results are shown in Figs.
\ref{3cond} and \ref{fdist05}.
The initial $f_0$ in Fig. \ref{3cond} is close to a Maxwellian and the
subsequent $f_t$ are also close to a Maxwellian with time dependent
temperatures.  In Fig. \ref{fdist05} the initial $f_0$ is one in which all
the particles have the same speed.  The evolution of $f_t$ towards a
Maxwellian is clearly visible.

{\bf Binary Mixtures}: We also considered the case of a binary system of hard point particles with
alternating masses $m_1$ and $m_2$ \cite{2mass}.  We used as our macrovariable the total energy $E$
plus the empirical density $f(v) dv$ of particles with velocities in some interval $dv$
(independent of the species) with uniform positional densities $n_1$ and $n_2$ in a box of length
$L$.  The entropy $S_{(2)}(f,E)$ of this system can be written as a sum:
$$
S_{(2)}(f,E) = S_{\rm gas}(f_1) + S_{\rm gas}(f_2), \quad E = E_1 + E_2
$$
The maximum of $S_{(2)}(f,E)$ is obtained for $f$ a sum of Maxwellians
$f_1$ and $f_2$ with the same temperature $T$ determined by the total
energy $E = (n_1 + n_2) {1 \over 2} kT L$ .

  Starting with an initial
microstate for which $S_{\rm gas}(f_1) > L s_{\rm eq}(n_1,T_1)$, ~~ $S_{\rm gas}(f_2) < L s_{\rm
eq}(n_2,T_2)$, we then observed in the simulations that $S_{\rm gas}(f_1)$ decrease while
$S_{(2)}(f,E)$ increase during the evolution towards the equilibrium distributions.
\begin{figure}
\centerline{\includegraphics[width=0.5 \textwidth]{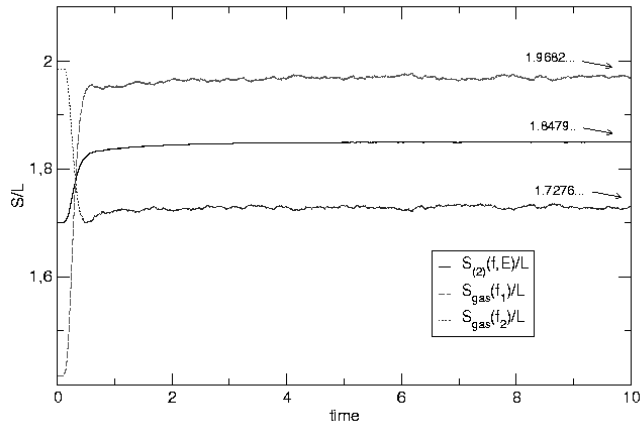}} \caption{Evolution of the
nonequilibrium entropy $S_{(2)}[f,E]/L$ for a one dimensional system with $N=10^5$ particles with
alternating masses $m_1=1$ and $m_2=(1+\sqrt{5})/2$. Initially the particles with mass $1$ ($2$)
have a maxwellian velocity distribution with temperature $T_1=1$ ($T_2=5$). $S_{\rm gas}(f_i)$ is
the partial entropy for the $i$-specie.}\label{twogaus}
\end{figure}

{\bf Concluding remarks}:  We have confirmed via computer simulations the
monotone increase of the Boltzmann entropy (log of phase space volume)
for a dense fluid not in local equilibrium whose macrostate is
specified by the empirical density $f$ and energy $E$.  Similar
results were obtained for a binary system of hard points in $d=1$.  The
simulations also show an apparent deterministic evolution of $f_t$ for
such systems.

{\bf Acknowledgements:} P.L.G  acknowledges the support of MEC (Spain), the MCYT-FEDER project
BFM2001-2841. The work of JLL was supported by NSF Grant DMR 01-279-26, and by AFOSR Grant AF
49620-01-1-0154. J.L.L. would like to thank O. Penrose for useful discussions.

\end{document}